# In memoriam: Dmitri Ivanenko (1904 – 1994)

## In honor of the 110[th] Year Anniversary

(G. Sardanashvily, *Science Newsletter*, Issue 1 (2014) 16)

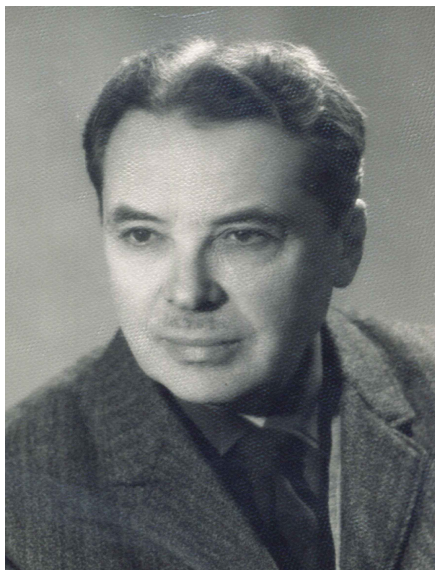

**Dmitri Ivanenko** (29.07.1904 – 30.12.1994), professor of Moscow State University**,** was one of the great theoreticians of XX century, an author of the proton-neutron model of atomic nucleus (1932).

D. Ivanenko was born on July 29, 1904 in Poltava (Russian Empire), where he began his creative path as a school teacher of physics. In 1923 Ivanenko entered Petrograd University. In 1926, while still a student, he wrote first scientific works with his friends George Gamov and Lev Landau (Nobel Laureate in 1962). After graduating the university, from 1927 to 1930 D. Ivanenko was a scholarship student and a researcher scientist at the Physical Mathematical Institute of Academy of Sciences of USSR. During these years he collaborated with Vladimir Fok and Viktor Ambartsumian, later to become famous.

In 1929 – 31, Dmitri Ivanenko worked at the Kharkiv Institute of Physics and Technology, being the first director of its theoretical division; Lev Landau followed him in 1932 – 37. Paskual Jordan, Victor Wieskopf, Felix Bloch (Nobel laureate in 1952) and Paul Dirac (Nobel Laureate in 1933) visited D. Ivanenko in Kharkiv. In Kharkiv, Ivanenko organized the 1[st] Soviet theoretical conference (1929) and the first soviet journal "*Physikalische Zeitschrift der Sowjetunion*" in foreign language (1932).

After returning to Leningrad at the Ioffe Physical-Technical Institute, D. Ivanenko concentrated his interest to nuclear physics. In May 1932, Ivanenko published his famous proton-neutron model of the atomic nucleus in "Nature" [18], and two months later Werner Heisenberg (Nobel laureate in 1932) referred to his work.

> **The Neutron Hypothesis**
>
> Dr. J. Chadwick's explanation [1] of the mysterious beryllium radiation is very attractive to theoretical physicists. Is it not possible to admit that neutrons play also an important rôle in the building of nuclei, the nuclei electrons being *all* packed in α-particles or neutrons? The lack of a theory of nuclei makes, of course, this assumption rather uncertain, but perhaps it sounds not so improbable if we remember that the nuclei electrons profoundly change their properties when entering into the nuclei, and lose, so to say, their individuality, for example, their spin and magnetic moment.
>
> The chief point of interest is how far the neutrons can be considered as elementary particles (something like protons or electrons). It is easy to calculate the number of α-particles, protons, and neutrons for a given nucleus, and form in this way an idea about the momentum of nucleus (assuming for the neutron a moment $\frac{1}{2}$). It is curious that beryllium nuclei do not possess free protons but only α-particles and neutrons.
>
> D. Iwanenko.
> Physico-Technical Institute,
>     Leningrad, April 21.
>
> NATURE, 129, 312, Feb. 27, 1932.

**Iwanenko D., The neutron hypothesis, Nature, v.129, N 3265, p.798, 1932**

In August 1932, D. Ivanenko and E. Gapon proposed the pioneer nuclear shell model describing the energy level arrangement of protons and neutrons in the nucleus in terms of energy levels [22]. Later this model was developed by Eugene Paul Wigner, Maria Goeppert-Mayer and J. Hans D. Jensen who shared the 1963 Nobel Prize for their contributions.

Ivanenko's success pushed forward the nuclear physics in the USSR. In 1933 on the initiative of Dmitri Ivanenko and Igor Kurchatov, the 1[st] Soviet nuclear conference was organized. Paul Dirac, Frédéric Joliot-Curie (Nobel laureate in 1935), Fransis Perrin, Ftanko Rasetti, Victor Wieskopf et al participated in this Conference.

The realization of Ivanenko's far-reaching plans and hopes was interrupted, however. In 1935 he was arrested in connection with the Sergey Kirov affair. Exile to Tomsk followed. D. Ivanenko was a professor at Tomsk and Sverdlovsk Universitie until the beginning of the World War II. From 1943 and until the last days of his life, he was closely associated with the Physics Faculty of M.V. Lomonosov Moscow State University.

Dmitri Ivanenko made the fundamental contribution to many areas of nuclear physics, field theory and gravitation theory.

In 1928, Ivanenko and Landau developed the theory of fermions as skew-symmetric tensors in contrast with the Dirac spinor model [4]. Their theory, widely known as the Ivanenko -- Landau – Kahler theory, is not equivalent to Dirac's one in the presence of a gravitational field, and only it describes fermions in contemporary lattice field theory.

In 1929, Ivanenko and Fock generalized the Dirac equation and described parallel displacement of spinors in a curved space-time (the famous Fock – Ivanenko coefficients) [9]. Nobel laureate Abdus Salam called it the first gauge field theory.

In 1930, Ambartsumian and Ivanenko suggested the hypothesis of creation and annihilation of massive particles which became the corner stone of contemporary quantum field theory [16].

In 1934 Dmitri Ivanenko and Igor Tamm (Nobel Laureate in 1958) suggested the first non-phenomenological theory of paired electron-neutrinor nuclear forces [24]. They made the significant assumption that interaction can be undergone by an exchange of massive particles. Based on their model, Nobel laureate Hideki Yukawa developed his meson theory.

In 1938, Ivanenko proposed a non-linear generalization of Dirac's equation. Based on this generalization, W. Heisenberg and he developed the unified nonlinear field theory in $50^{th}$ [69].

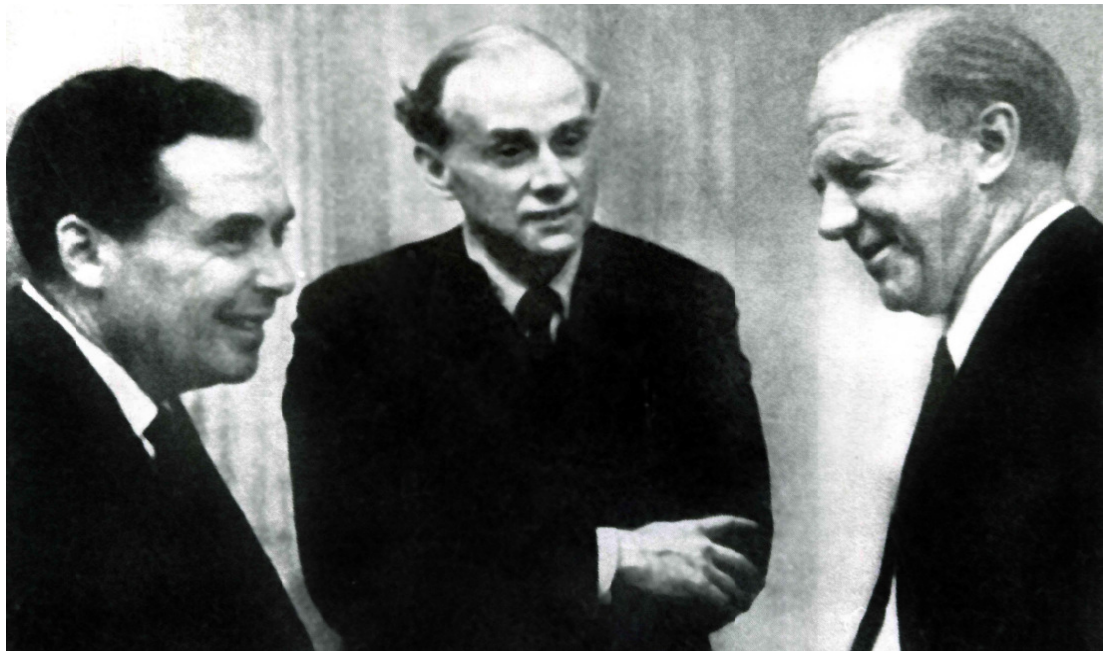

**D. Ivanenko, P.A.M. Dirac and W. Heisenberg (Berlin, 1958)**

In 1944, Dmitri Ivanenko and Isaak Pomeranchuk predicted the phenomenon of synchrotron radiation given off by relativistic electrons in a betatron [39]. This radiation was soon discovered by American experimenters D. Bluitt (1946) and H. Pollock (1947). Synchrotron radiation possesses a number of very particular properties which provide its wide applications. In particular, neutron stars also are sources of this type radiation. Classical theory of synchrotron radiation was developed by Dmitri Ivanenko in collaboration with Arseny. Sokolov in 1948, and independently by Julian Schwinger (Nobel Laureate in 1965). For their work on synchrotron radiation, D. Ivanenko, A. Sokolov and I. Pomeranchuk were awarded the Stalin Prize in 1950.

PHYSICAL REVIEW     VOLUME 65, NUMBERS 11 AND 12     JUNE 1 AND 15, 1944

## Letter to the Editor

PROMPT *publication of brief reports of important discoveries in physics may be secured by addressing them to this department. The closing date for this department is the third of the month. Because of the late closing date for the section no proof can be shown to authors. The Board of Editors does not hold itself responsible for the opinions expressed by the correspondents. Communications should not in general exceed 600 words in length.*

### On the Maximal Energy Attainable in a Betatron

D. IWANENKO AND I. POMERANCHUK
*Physical Institute of the Moscow State University, Moscow, and Physico-Technical Institute of the Academy of Sciences of the U.S.S.R., Leningrad, U.S.S.R.*
May 18, 1944

BY means of a recently constructed induction accelerator-betatron, Kerst succeeded in obtaining electrons up to 20 Mev.[1] The principle of operation of the betatron is the acceleration of electrons by a tangential electric field produced by a changing magnetic flux, which is connected with the magnetic field keeping electrons on the orbit by a simple relation. In contrast to a cyclotron, whose applicability is essentially limited to the non-relativistic region on the ground of defocusing of orbits due to the change of mass at high energies, there is no such limitation for the betatron.

We may point out, however, that quite another circumstance would lead as well to the existence of a limitation for maximal energy attainable in a betatron. This is the radiation of electrons in the magnetic field. Indeed, electrons moving in a magnetic field will be accelerated and must radiate in accordance with the classical electrodynamics. One can easily see that quantum effects do not play here any important role as the dimension of the orbit is very great. As was shown by one of us[2] an electron moving in a magnetic field H radiates per unit of path the energy

$$-(dE/dX) = 2/3(e^2/mc^2)^2(E/mc^2)^2[(V/c)H]^2 \quad (1)$$

where $e$ is the charge, $m$ the mass, $V$ the velocity, and $E$ the energy of the electron; $E$ is assumed much greater than $mc^2$.

In the betatron $V$ is normal to $H$ and practically for the whole path equal to $c$. Then we have

$$-(dE/dX) = 2/3(e^2/mc^2)^2(EH/mc^2)^2. \quad (2)$$

The limiting value of energy $E_0$ is to be determined from the condition that the radiated energy (2) will be equal to energy gained by the electron in the electric field produced by magnetic flux per unit of path:[3]

$$\frac{2}{3}e^2\left(\frac{E_0H}{mc^2}\right)^2 = \frac{e|d\phi/dt|}{2\pi R_0 c} = \frac{e}{c}R_0|\dot{H}| \quad (3)$$

$$\dot{H} = dH/dt \quad r_0 = e^2/mc^2.$$

Here $R_0$ is the radius of the orbit, $\phi$ is the induction flux.[1] Hence:

$$\frac{E_0}{mc^2} = \left(\frac{3eR_0}{2r_0 c}\frac{\dot{H}}{H^2}\right)^{\frac{1}{2}}. \quad (4)$$

Taking for $H$ and $E$ the values now being in use we get $E_0 \approx 5 \times 10^8$ ev, which is only five times as great as the energy which one expects to obtain in the betatron now under construction. From (4) one sees that $E_0$ is inversely proportional to the magnetic field applied and proportional to the square root of energy gained in the rotation electric field per unit of path. All this requires the using of smaller $H$ or of higher frequencies with the purpose of getting higher limiting values of $E_0$.

The radiative dissipation of energy of electrons moving in a magnetic field must be also of importance for the discussion of the focusing of the electronic beam, as the energy of particles being accelerated will grow more slowly with the growth of $H$ if the radiation is taken into account. This latter question may deserve a separate discussion.

[1] D. W. Kerst, Phys. Rev. 61, 93 (1942).
[2] I. Pomeranchuk, J. Phys. 2, 65 (1940).
[3] D. W. Kerst and R. Serber, Phys. Rev. 60, 53 (1941).

**Iwanenko D., Pomeranchuk I., On the maximal energy attainable in betatron, Physical Reviews, v.65, p.343, 1944**

Two of D. Ivanenko's and A. Sokolov's monographs "*Classical Field Theory*" and "*Quantum Field Theory*" were published at the beginning of the 50th. "Classical field theory" was the first contemporary book on field theory where, for instance, the technique of generalized functions was applied. Nobel laureate Ilya Prigogine referred to it as his text-book.

In 1956, D. Ivanenko developed the theory of hypernuclei discovered by Marian Danysz and Jerzy Pniewski in 1952.

At the beginning of the 1960's, D. Ivanenko did intensive scientific and organizational work on the development and coordination of gravitation research in the USSR. In 1961, on his initiative the 1st Soviet gravitation conference was organized. D. Ivanenko was the organizer of Soviet Gravitation Commission, which lasted until the 1980's. He was a member of the International gravitation Committee since its founding in 1959.

In the 70 – 80th, D. Ivanenko was concentrated on gravitation theory. He developed different generalizations of Einstein's General Relativity, including gravity with torsion, the hypothesis of quark stars [63] and gauge gravitation theory [79]. In 1985, D. Ivanenko and his collaborators published two monographs "*Gravitation*" and "*Gauge Gravitation Theory*".

Theoretical physics in the USSR has been enormously influenced by the seminar on theoretical physics organized by D. D. Ivanenko in 1944 that has continued to meet for 50 years under his guidance at the Physics Faculty of Moscow State University. The distinguishing characteristic of Ivanenko's seminar was the breadth of its grasp of the problems of theoretical physics and its discussion of the links between its various divisions, for example, gravitation theory and elementary particle physics. The most prominent physicists in the world participated in the seminar: Niels and Aage Bohr, Paul Dirac, Hideki Yukawa, Julian Schwinger, Abdus Salam, Ilya Prigogine, Samuel Ting, Paskual Jordan, Tullio Regge, John Wheeler, Roger Penrose et al.

The scientific style of Dmitri Ivanenko was characterized by great interest in ideas of frontiers in science where these ideas were based on strong mathematical methods or experiment.

It should be noted that seven Nobel Laureates: P.A.M. Dirac, H. Yukawa, N.Bohr, I. Prigogine, S. Ting, M. Gell-Mann, G. 't Hooft wrote their famous inscriptions with a chalk on the walls of Ivanenko's office in Moscow State University.

### Selected publications of D. Ivanenko

1. Gamov G., Iwanenko D., Zur wellentheorie der materie, Zeitschrift für Physik, Bd.39, s.865-868, 1926.
2. Iwanenko D., Landau L., Zur albeitung der Klein-Fockschen gleichung, Zeitschrift für Physik, Bd.40, s.161-162, 1927.
3. Iwanenko D., Landau L., Bemerkung über quantenstatistik, Zeitschrift für Physik, Bd.42, s.562, 1927.
4. Iwanenko D., Landau L., Zur theorie des magnetischen electrons. I, Zeitschrift für Physik, Bd.48, s.340-348, 1928.
5. Ivanenko D., Über eine verallgemeinerung der geometrie, welche in der quantenmechanik nützlich sein kann, ДАН СССР, N4, c.73-78, 1929.
6. Iwanenko D., Deux remarques sur l'equation de Dirac, Compt. Rend. Acad Sci. Paris, v.188, p.616-618, 1929.
7. Fock V., Iwanenko D., Über eine mögliche geometriche deutung der relativistichen quantentheorie, Zeitschrift für Physik, Bd.54, s.798-802, 1929.
8. Iwanenko D., Bemerkung über quantengeschwindligkeit, Zeitschrift für Physik, Bd.55, s.141-144, 1929.
9. Fock V., Iwanenko D., Géometrie quantique linéaire et déplacement paralléle, Compt. Rend. Acad Sci. Paris, v.188, p.1470-1472, 1929.
10. Fock V., Iwanenko D., Zur quantengeometrie, Phys. Z., Bd.30, s.648, 1929.
11. Fock V., Iwanenko D., Quantum geometry, Nature, v.123, p.838, 1929.

*Gennadi Sardanashvily*

*Department of Theoretical Physics, Moscow State University*